\documentclass[runningheads]{llncs}
\usepackage{graphicx}

\usepackage{cite}
\usepackage{amssymb,amsfonts}
\usepackage{algorithm}
\usepackage{algorithmic}
\usepackage{graphics}
\usepackage{array,booktabs}
\usepackage{graphicx}
\usepackage{textcomp}
\usepackage{xcolor}
\usepackage{tabularx}
\usepackage[intlimits]{amsmath}
\usepackage{multirow}
\usepackage{adjustbox}
\usepackage[utf8]{inputenc}
\usepackage{tikz}
\usetikzlibrary{positioning}
\usetikzlibrary{shapes,arrows,arrows,positioning,fit}

\def\BibTeX{{\rm B\kern-.05em{\sc i\kern-.025em b}\kern-.08em
    T\kern-.1667em\lower.7ex\hbox{E}\kern-.125emX}}
\begin{document}
\title{A Non-Intrusive Machine Learning Solution for Malware Detection and Data Theft Classification in Smartphones}

\author{Sai Vishwanath Venkatesh \inst{1}\orcidID{0000-0001-6568-6259} \and Prasanna Kumaran D
\inst{2} \and Joish J Bosco 
\inst{1} \and Pravin Kumaar R
\inst{1} \and Vineeth Vijayaraghavan \inst{1}}
\authorrunning{SV. Venkatesh et al.}
\institute{ Solarillion Foundation, Chennai, India \\
\email{\{saivishwanathv,pravin.kumaar99,vineethv\}@ieee.org}\\ \email{joishbosco99@gmail.com} \and
SSN College of Engineering, Chennai, India \\
\email{prasannakumaran18110@cse.ssn.edu.in}}
\titlerunning{A Non-Intrusive Solution for Data Theft Classification in Smartphones}
\maketitle              
\begin{abstract}
Smartphones contain information that is more sensitive and personal than those found on computers and laptops. With an increase in the versatility of smartphone functionality, more  data has become vulnerable and exposed to attackers. Successful mobile malware attacks could steal a user's location, photos, or even banking information. Due to a lack of post-attack strategies firms also risk going out of business due to data theft. Thus, there is a need besides just detecting malware intrusion in smartphones but to also identify the data that has been stolen to assess, aid in recovery and prevent future attacks. In this paper, we propose an accessible, non-intrusive machine learning solution to not only detect malware intrusion but also identify the type of data stolen for any app under supervision. We do this with android usage data obtained by utilising publicly available data collection framework--\textit{SherLock}. We test the performance of our architecture for multiple users on real-world data collected using the same framework. Our architecture exhibits less than 9\% inaccuracy in detecting malware and can classify with 83\% certainty on the type of data that is being stolen.

\keywords{ High-dimensional data  \and Machine Learning \and Cybersecurity \and Data Mining}
\end{abstract}

\section{Introduction}

Currently, Android has more than 1.6 billion active users, which accounts for more than 70\% of the global market share of mobile operating systems. As a result, the application market for android is flooded with apps. We define \textit{malicious apps} or \textit{malware} as Android applications that present itself to the user as benign, but secretly steals user information in the background. Although the Android application store (Google Play) verifies apps for malicious intent upon release, it does not vehemently track updates from these verified apps and can not account for third-party apps downloaded independently by the user.  A report released in 2020 by McAfee Advanced Threat Research and Mobile Malware Research \cite{mcAfee2020} suggests that malware developers roll out malware through verified apps in Google Play as updates to shield themselves from preliminary verification. Undetected malware attacks can steal sensitive and organization-crippling information from users such as photos, documents and browsing data. Data breaches are extremely disastrous for small and midsize firms and businesses. 
A report by the U.S. Securities and Exchange Commission\cite{usSEC2015} states that 60\% of small firms can not recuperate from data breaches and go out of business within 6 months. The IBM "Cost of a Data Breach Report 2020"\cite{ibm2020} suggests companies to establish an incident response (IR) plan to determine the damage done by the breach and contain it as soon as possible. It goes on to state that companies with an IR plan save an average of 2\$ million in the event of a data breach. Furthermore, the report projects an increase in the costs of data breaches due to the COVID-19 pandemic and the increase in digital reliability. This calls for a need to not only detect malicious attacks but to also identify the stolen data to assess the damage, strategically recover and prevent future attacks. Performing this can help in understanding malware trends and aid in malware prevention research.

We propose a novel two-stage machine learning approach to detect malicious attacks for any app under supervision and identify the data stolen by the attack to aid in assessment and recovery. 

The course of this paper is as follows: Section \ref{relatedWorks} discusses the research works carried out with relevance to malware detection. In Section \ref{datasetSection}, we describe the dataset used in our study extensively. We elucidate the steps taken to make the data computationally feasible in Sections  \ref{data-pre} and \ref{featureSelectionSection}. Later, in Section \ref{modelingAndExperimentation} and \ref{EvaluationMetricsSection}
we outline our model architecture and describe the parameters of its evaluation.  Finally, in Section \ref{ResultsandDiscussions} we report and discusses our findings.

\section{Related Work}\label{relatedWorks}

Mobile malware detection has been an active and broad area of research for the past several years. Static analysis was one of the first major mobile malware detection approaches proposed \cite{ShmmidtExecutables,demystified}. Here, the source code of the target malware is analyzed to identify semantic signatures. Although static analysis can detect malware even before running the app, static analysis systems fail when the malware uses obfuscation techniques such as code encryption and repackaging. Dynamic analysis techniques \cite{taintdroid, sandbox} address code obfuscation and encryption in malware detection by executing the source code of the application in an isolated environment to analyze runtime characteristics based on frequency. However, this proves to be a bottleneck in dynamic analysis systems as isolated, lab-like noiseless data is hard to achieve and be implemented in a real-world setting. Static and dynamic methods additionally require super-user (root) access  since they require source code to be implemented. Furthermore, Moser et al. \cite{limitsofstatic} suggests that the rate of developing rule-based solutions can not match the fast rate of new malware released to the world.  thus, these solutions will fail to perform for new malware since they are rule-based and specific solutions.

Machine learning approaches were introduced to swiftly aid in detecting new malware as they are released. Notable works using these approaches include \cite{trafficAV,highImbalanceNetwork,arora,unkownMalware} that outperforms static and dynamic methods by modelling network usage for detection. 
[7] used various anomaly detection methods to detect malware using system and network data collected. 
Ronen et. al\cite{microsoft2018} and related works go on to detect and classify the family of the detected malware by analysing dalvik bytecode from android devices. However, these works fail to address the security risk for any end user to obtain bytecode. This exposes the phone to further vulnerabilities due to the need for root access. There is a need for non-intrusive malware detection systems based on low privilege information such as usage statistics. This would allow easier user applicability and ensure better security over super-user vulnerabilities.

We propose modeling malware on usage statistics data and we consider one of the largest and most granular dataset for mobile sensor and software  sampling - Sherlock Dataset 
As a result of the dataset's versatility it is suitable for a multitude of use-cases. Since it does not require root access to probe its data it is safe and reproducible for malware detection. 
Zheng et al.\cite{mobileAppAndMalwareClassification} explored usage patterns and the relationship between mobile usage and the state (benign/malicious) of the application for this data.
Wassermann et al.\cite{bigmomal} used low-level system features from this dataset with sampling techniques to deal with an inherent class imbalance and detect malicious actions performed on a smartphone.

Although, current research tackles malware detection extensively they fail to address data theft classification to aid damage assessment and recovery from data breaches. 

We consider using the SherLock dataset  to develop a machine learning malware detection pipeline that would detect if any given app is malicious and identify the data it attempts to steal if detected. We aim to utilize network traffic data, local and global system features to implement this solution.
\vspace{-2.5mm}
\section{Dataset}\label{datasetSection}
For our experiments we used the SherLock dataset\cite{Sherlock}. The SherLock Dataset, spanning over 10 billion records, involving over 50 volunteers is the result of a real-world data collection experiment to obtain low-level Android usage data alongside emulated malware. Such statistics do not require root access, therefore making any solution developed on the dataset more secure under real-world circumstances since rooting exposes a mobile phone to further vulnerabilities. 

The experiment introduces two data collection agents to the mobile phones provided to the volunteers -- \textit{Sherlock} and \textit{Moriarty}. \textit{Moriarty} emulates malicious actions on the volunteer's mobile phones randomly through the course of the experiment generating distinct labels between malicious and benign actions. Meanwhile, \textit{Sherlock} logs usage attributes and statistics in the background.

\begin{table}[]
\caption{Categories of Data Theft}
\vspace*{-1.5mm}
\label{dataTheft}
\resizebox*{\columnwidth}{!}{

\begin{tabular}{|c|c|}
\hline
\textbf{Malware Service Type} & \textbf{Target information}                            \\ \hline
Contacts                      & Phonebook data                                         \\ \hline
GPS                           & User coordinates (latitude and longitude)              \\ \hline
URL                           & Web address of every page visited by the user recently \\ \hline
Audio Records                 & Audio records collected during the session             \\ \hline
Contacts                      & Names and Phone numbers                                \\ \hline
BrowserInfo                   & Account details, bookmarks and browser history         \\ \hline
Photos                        & Images from gallery                                    \\ \hline
\end{tabular}
 }
\vspace{-4mm}
\end{table}

\subsection{Sherlock Data Collection Agent}
One of the ways Sherlock logs phone attributes is through \textit{Pull Probes} which extract data periodically at a constant sampling rate. For our experiments we consider the most frequently sampled pull probe in Sherlock named \textit{T4}, which has a sampling rate of 5 seconds. T4 probes Global System Features as well as Local Application Features.

\vspace{2mm}
\textbf{Global System Features (GSF)} These features pertain to attributes with a global scope in the Android system such as network traffic, CPU and memory utilization, IO interrupts and WiFi related data. There are a total of 128 Global System Features.
\vspace{2mm}

\textbf{Local Application Features (LAF)} Alongside Global System Features, Linux-level data \cite{linuxMan} for every running application is sampled. This includes process specific features such as the scheduling priority, number of bytes transferred, number of threads and kernel level features used by an application at the time instant. There are a total of 56 Local Application Features.

Local Application Features used in context with Global System Features provides a rich feature set to determine if a given app exhibits malicious behaviour.

\subsection{Moriarty Malicious Agent}
Moriarty presents itself to the user as a benign application, such as a Game or a Browser depending on the version of the app but covertly performs malicious actions. The malware emulated by each version is dissimilar to its precursor and targets different vulnerabilities in each version as illustrated in Table \ref{dataTheft}. The malware used by Moriarty are behavioural copies of malware found in the real-world.

The app contains labels indicating whether an action executed is benign or malicious. Furthermore, the details of the malicious actions such as the type of data stolen, number of bytes transmitted and time taken to transfer the stolen information are logged along with the labels. To collect sufficient information for the experiment the volunteers were reminded to use the Moriarty app if they have not used it for a couple of days.

For the experiment we have considered a computationally feasible subset of the SherLock dataset. It consists of data collected during the first quarter of 2016, with over 300 million records, spanning across 5 users.

\section{Data Pre-processing}
\label{data-pre}
We aim to enable efficient data merging between Local Application Features (\textit{LAF}) and Global System Features (\textit{GSF}). Let \textit{g} and \textit{n} denote the number of GSF and LAF. Assuming there are \textit{m} apps running at the same time, each Global System Feature would correspond to multiple LAF at that instant of time. The vector space of application data (LAF at time \textit{t}), denoted by $\Omega_t$ for any time instant \textit{t} is represented in Equation \eqref{omega}.
\begin{align}
    \Omega_{t} &= \begin{Bmatrix}
    \omega_{11}       & \omega_{12}  & \dots & \omega_{1n} \\
    \omega_{21}       & \omega_{22}  & \dots & \omega_{2n} \\
    \vdots{}     &\vdots{} & \vdots{}& \vdots{}\\
    \omega_{m1}       & \omega_{m2}  & \dots & \omega_{mn} \label{omega}
    \end{Bmatrix}
\end{align}

Consequently, if a relational join operation between GSF and LAF was performed it would lead to the generation of GSF duplicates for every running application with a shape of ($m,\, g + n$). The size of this data denoted by $S_{np}$ is \textit{$m *(g + n)$} memory units.
With the dataset spanning over 300 million records, it becomes essential to reduce memory consumption to expedite the data handling and modeling process. Therefore to overcome duplicates, $\Omega_t$ is transformed into a row vector of shape ($1,\, m * n$) by performing \textit{PIVOT} operation represented in Equation \eqref{pivot}, thus obtaining a functional dependency with time.
\begin{equation}
    PIVOT(\Omega_{t}):= \{\omega_{ij} \mid i \in{M}\ and \, j \in{N}\} \label{pivot}
\end{equation}

\begin{center}
\small{$M$ = Set of all applications on the device\\
$N$ = Set of local application features
}\end{center}

As a result of using $PIVOT(\Omega_t)$ to merge with GSF as opposed to using $\Omega_t$ we obtain  a shape of $(1,\,g + m * n)$ and size of this data denoted by $S_{p}$ is $1 *(g + m * n)$. The size comparison of the data obtained from merging GSF with and without pivot operation is illustrated in Equations \eqref{inequality}.
Thus, reducing the overall throughput as the number of applications increase.
\begin{equation}
\begin{split}
     & \;g + m * n << g * m + m * n \\
     &\implies \;\;S_{p} << S_{np} \label{inequality}
\end{split}
\end{equation}
For the first quarter of 2016 in SherLock, $\,g=128$ and $n=56$ with an average of $m=55$ running at any given time. 

We observed that {$S_{np}$} / {$S_{p}$} was 3.2 indicating that the pivot operation was effective in reducing the size of the merged data. We obtain a dataset with 14,234 features and 5.81 million records on merging this data with Moriarty labels.


%

\section{Feature Selection}\label{featureSelectionSection}
We strive to reduce the dataset to its most informative features for smooth and utilitarian processing. On closer inspection of the 14,234 features, we discovered 12,726 features to have more than 70\% of null values in them and we obtain 1508 features as a result of their removal. However, this remains significantly large for us to process, considering that we have 5.8 million records. 

To further reduce the feature set, we pursue a feature selection method that ensures relevance towards our objective -- malware detection and target classification. We considered LightGBM \cite{lightgbm} as it has proven to be fast and scalable especially when implemented on high dimensional datasets \cite{biology}. Using this technique we reduce our feature space to 150 and 100 important features for malicious detection and target classification respectively 

With the features reduced to less then 15\% of 1508 features, we can now implement stepwise forward selection \cite{forwardFeatureSelection} -- an iterative method to determine the least number of features required to obtain any given model's best performance. Using stepwise forward selection we reduce the features required to detect malware to 10 features and the features required to determine the data targeted by malware to 16 features.

As a result of our feature selection approach the feature set is reduced to approximately 0.1\%  of the original feature set. Table \ref{SelectedFeatures} lists the most important features that were considered for modeling. 

\begin{table}[]
\centering
\caption{Features Selected for Proposed Architecture}
\vspace*{-1.5mm}
\label{SelectedFeatures}
\resizebox{\columnwidth}{!}{
\begin{tabular}{|c|c|c|}
\hline
\textbf{Model Stage} & \textbf{GSF}                                                                                                          & \textbf{LAF}                                                                                                                                                                                                                                                                                                                                                                                                 \\ \hline
Malware Detector     & \begin{tabular}[c]{@{}c@{}}totalmemory\_used\_size, \\ totalmemory\_freesize, \\ traffic\_totalrxpackets\end{tabular} & \begin{tabular}[c]{@{}c@{}}dalvikprivatedirty\_Moriarty, \\ dalvikprivatedirty\_WhatsApp, \\ dalvikpss\_Samsung Push Service, \\ otherpss\_SherLock, rss\_SherLock, \\ uidrxbytes\_Moriarty, \\ num\_threads\_SherLock\end{tabular}                                                                                                                                                                          \\ \hline
Target Classifier    & --                                                                                                                    & \begin{tabular}[c]{@{}c@{}}utime\_SherLock, rss\_SherLock, \\ utime\_Moriarty, stime\_Moriarty, \\ importance\_SherLock, lru\_SherLock, \\ dalvikprivatedirty\_SherLock, \\ vsize\_Hangouts, num\_threads\_Moriarty, \\ rss\_Hangouts, otherpss\_Hangouts, \\ dalvikpss\_Hangouts, \\ num\_threads\_SherLock, \\ utime\_Unified Daemon, \\ otherprivatedirty\_Context Service, \\ vsize\_Chrome\end{tabular} \\ \hline
\end{tabular}
}
\vspace{-10mm}
\end{table}

\section{Modeling and Experimentation}\label{modelingAndExperimentation}



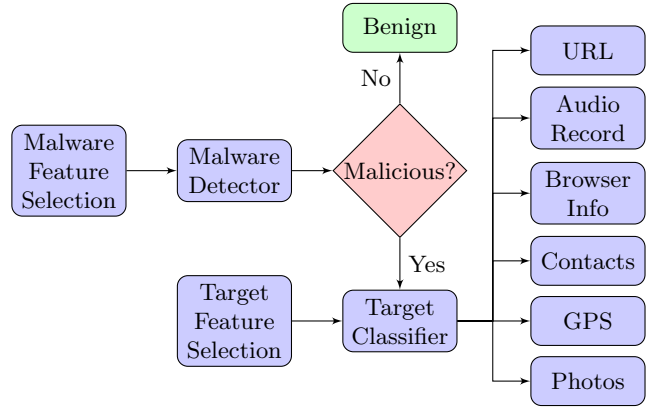
\begin{figure}
\centering
 	\tikzstyle{decision} = [diamond, draw, fill=blue!20, 
 	text width=5.5em, text badly centered, inner sep=0pt]
 	\tikzstyle{block} = [rectangle, draw, fill=blue!20, 
 	text width=7em, text centered, rounded corners, minimum height=2em]
 	\tikzstyle{line} = [draw, -latex']
 	\tikzstyle{cloud} = [draw, ellipse,fill=red!20,
 	minimum height=4em]
 	\tikzstyle{decisionn} = [diamond, draw, fill=red!20, 
 	text width=4.8em, text badly centered, node distance=4cm, inner sep=0pt]
 	\tikzstyle{blockk} = [rectangle, draw, fill=blue!20, 
 	text width=4em, text centered, rounded corners, minimum height=2em]
 	\tikzstyle{blockkB} = [rectangle, draw, fill=green!20, 
 	text width=4em, text centered, rounded corners, minimum height=2em]
 	\tikzstyle{blockkk} = [rectangle, draw, fill=red!20, 
 	text width=6em, text centered, rounded corners, minimum height=2em]

 	
 	\begin{tikzpicture}[node distance = 2cm, auto]
    \node [blockk,] (init) {Target Classifier};
    \node [coordinate,  right=0.001cm of init] (rightt) {};
 	\node [decisionn, above of=init, node distance=2cm] (node0) {Malicious?};
 	\node [blockk, left of=node0, node distance=2.2cm] (node1) {Malware Detector};
 	\node [blockk, left of=node1, node distance=2.2cm] (node2) {Malware Feature Selection};
 	\node [blockk, left of=init, node distance=2.2cm] (node10) {Target Feature Selection};
\node [blockkB, above of=node0, node distance=1.9cm] (node11) {Benign};
    
 	\node [blockk, right of=init, node distance=2.5cm] (intt) {GPS};
 	
 	\node [blockk, above of=intt, node distance= .8cm] (node5) {Contacts};
 	\node [coordinate,  left=0.5cm of node5] (q) {};
 	
 	\node [blockk, above of=node5, node distance= .9cm] (node6) {Browser Info};
 	\node [coordinate,  left=0.5cm of node6] (w) {};
 	
 	\node [blockk, below of=intt, node distance= .8cm] (node7) {Photos};
 	\node [coordinate,  left=0.5cm of node7] (e) {};
 	
 	\node [blockk, above of=node6, node distance= 1cm] (node9) {Audio Record};
 	\node [coordinate,  left=0.5cm of node9] (t) {};
 	
 	\node [blockk, above of=node9, node distance= .9cm] (node8) {URL};
 	\node [coordinate,  left=0.5cm of node8] (r) {};

 	\path [line] (init) -- (intt);
 	\path [line] (node10) -- (init);
 	\path [line] (init.east) |-(rightt)-|(q)-- (node5);
 	\path [line] (init.east) |-(rightt)-|(w)-- (node6);
 	\path [line] (init.east) |-(rightt)-|(e)-- (node7);
 	\path [line] (init.east) |-(rightt)-|(r)-- (node8);
 	\path [line] (init.east) |-(rightt)-|(t)-- (node9);
 	\path [line] (node0) -- node {Yes} (init);
 	\path [line] (node0) -- node {No} (node11);
 	\path [line] (node1) -- (node0);
 	\path [line] (node2) -- (node1);
 	\end{tikzpicture}
 	
 	\caption{Two-stage Architecture}
 	\label{flowChart}
 \end{figure}

Knowing the kind of data the malware steals could be of more use during data breach assessment compared to just detecting the presence of a malicious action. We propose a two-stage architecture illustrated in Fig. \ref{flowChart} to classify data targeted by a positively detected malware. Our approach detects if a malicious action  occurs  in  the  first  stage  and  if  positively  detected, classifies the data targeted during the malicious action in the second stage.

\vspace{-10mm}
\subsection{Malware Detection}
We primarily consider supervised tree-based models (Extra trees, Random forest, Decision tree and XGBoost) for malicious detection since they have proven to be effective for the data in use \cite{mobileAppAndMalwareClassification,bigmomal,apache}.
Anomaly detection methods are suggested by Mirsky et al. \cite{Sherlock} due to the sparse frequency of malicious records observed in the data as compared to benign(1:90). We aim to identify if anomaly detection methods are effective as per prior assumption, therefore we consider a tree-based anomaly and outlier detection method--Isolation Forest.

\subsection{Target classification}
We pass the values detected as malicious in the first stage to further classify the data targeted in this stage. This is a multi--class classification problem to determine the type of data targeted by the malware as seen in Table \ref{dataTheft}. 
We consider models--Extra trees, XGBoost and K-nearest neighbours for this task.
\section{Evaluation Metrics}\label{EvaluationMetricsSection}



\subsection{Malware Detection}\label{malwareDetectionEvalMetrics}

In this paper, we propose using \textit{False Omission Rate} (FOR) and \textit{False Positive Rate} (FPR) to evaluate the performance of a malicious detector. Accuracy and true positive rates as considered by \cite{andromaly,trafficAV,arora,mobileAppAndMalwareClassification} are not ideal choice of metrics as they evaluate the model's performance using the true positive values of the majority class which are generally high for highly imbalanced data such as \textit{SherLock} and therefore compensate the impreciseness in classifying the minority class. 


We aim to reduce the number of instances where a malware is misclassifed as benign. Therefore we consider \textit{False Omission Rate} (FOR) and \textit{False Positive Rate} (FPR) to evaluate the performance of the malicious detector. 

\subsubsection{\textbf{FOR}} Illustrated in Equation (\ref{FOR}) this metric indicates the fraction of malicious actions that go undetected by a malicious classifier.
\begin{equation}
\footnotesize FOR ={\frac{Number\,of\, malicious\,\,\, records\,\,\, predicted\,\,\, benign}{Total\,\,\,malicious\,\,\, records}}\label{FOR}
\end{equation}
\subsubsection{\textbf{FPR}}
Illustrated in Equation \eqref{FPR} this metric indicates the fraction of benign records that are misclassified.
\begin{equation}
\footnotesize FPR ={\frac{Number\,of\, benign\,\,\, records\,\,\, predicted\,\,\, malicious}{ Total\,\,\,malicious\,\,\, predictions}} \label{FPR}
\end{equation}

Although each metric can be used individually, we propose using both FOR and FPR in conjunction to discover a detector with an overall good-fit for detecting presence of malawre. A lower FOR signifies the success of the first stage of our architecture (malware detection). Meanwhile, a lower FPR signifies a smaller error that will cascade to the next stage.  Ideally both FOR and FPR need to be minimised to improve performance in data classification isn the second stage of our proposed two stage architecture.

\subsection{Target Classification}

Target Classification is a multi-class classification task that involves predicting what kind of data has been stolen by the malware. The different types of malware stolen were given equal importance and hence equal weights were considered for all the classes. Therefore, the average F1-score is the metric of choice used to evaluate the model in this stage.

\begin{figure}
    \centering
    \includegraphics[width=\linewidth]{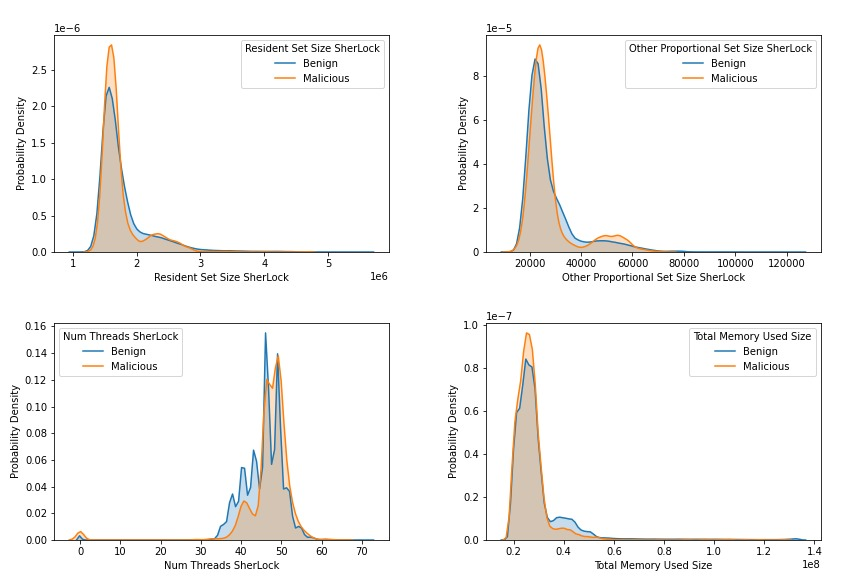}
    \caption{Density distributions of some of the most important features}
    \label{fig:isolation}
\vspace{-5mm}
\end{figure}

\begin{table}[]
\renewcommand{\arraystretch}{1.2}
\centering
\caption{Malware Detection Results}
\label{malwareDetectionResults}
\begin{tabular}{|c|c|c|}
\hline
\textbf{Classifier} & \textbf{False Omission Rate} & \textbf{False Positive Rate} \\ \hline
Decision tree & 0.063          & 0.222          \\ \hline
\textbf{Extra trees}    & \textbf{0.087} & \textbf{0.019} \\ \hline
Random forest           & 0.088          & 0.058                \\ \hline
XGBoost          & 0.646          & 0.296          \\ \hline
Isolation forest                & 0.793          & 0.976          \\ \hline
\end{tabular}
\vspace{-7mm}
\end{table}

\section{Results and Discussions}
\label{ResultsandDiscussions}

To evaluate the performance of our proposed architecture, we consider training and testing on all the users combined. Each user has been proportionally sampled (stratified) while splitting into 75\% for training and 25\% for testing.  
Since the proposed architecture consists of two stages, it cascades performance at each level. We are reporting the results at each stage for a deeper understanding of our performance.

\subsection{Malware Detection}

Tree-based classifiers prove to display superior performance for this task as illustrated in Table \ref{malwareDetectionResults}. This is due to their ability to capture discrete and categorical information more accurately.

However, contrary to prior assumption tree-based outlier detection method -- Isolation forest, fails to detect malware with a FOR of 0.79. On observing the density distributions of some of the most important features in Fig. \ref{fig:isolation} we discover an overlap between malicious and benign distributions. Anomaly detection methods are effective to identify outliers from distributions \cite{andromaly}. Since unsupervised and anomaly detection methods rely on the malware to exist outside benign distribution these methods may fail to detect malicious activity for this data.

With the least FOR of all the models considered (illustrated in Table \ref{malwareDetectionResults}), Decision tree and Extra trees are the best malware detectors with 6.3\% and 8.7\% FOR respectively. However, on closer inspection of the Decision tree detector we observe that it can only achieve this accuracy at the cost of 22.2\% FPR. Since this is not desirable for a performance cascading architecture as discussed in Section \ref{malwareDetectionEvalMetrics}, we use Extra trees to determine if an action is malicious before we classify its target in the next stage of our two-stage model.

\subsection{Progressive learning}
\begin{figure}
    \centering
     \includegraphics[scale = 0.6]{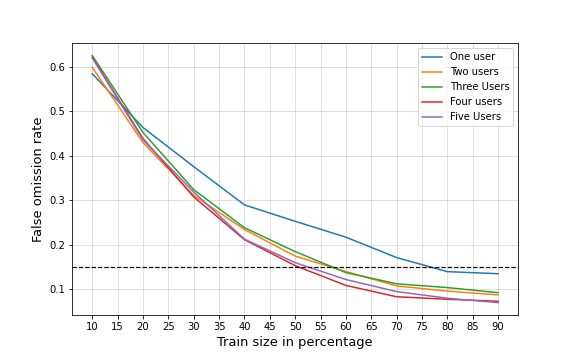}
    \caption{Progressive learning with increase in users}
    \label{progressiveLearning}
\vspace{-7mm}
\end{figure}

It is necessary for any user to be trained using the SherLock framework before the user can successfully monitor a newly installed app from the market. The time taken by each user to train the detector with SherLock would desirably need to be reduced which can be done by minimising the required train data for the detection task.
Our detector tackles this problem by combining all the users we have by performing stratified training and testing. Our detector exhibits the same accuracy with a decrease in train size as the number of users it has learned on increases. This is visualized in Fig. \ref{progressiveLearning} where we consider a threshold accuracy of 0.15 FOR to analyse the change in required train data for an Extra trees detector trained on 1-5 users. To achieve the threshold accuracy when our detector had trained only on a single user, the detector required atleast 76\% train data. However our detector reduces the percentage of train data required by each user as it learns from more users. When the model was trained on 5 users it required only 52.5\% of the train data to achieve the threshold FOR.



\newpage



\begin{table}
\label{targetDetectionTable}
\caption{Target detection results}
\centering
\begin{adjustbox}{max width=\textwidth}
\begin{tabular}{|c|c|c|c|c|c|c|c|} 
\hline
\multirow{2}{*}{\textbf{Models }} & \multirow{2}{*}{\textbf{Average~F1 Score }} & \multicolumn{6}{c|}{\textbf{Class-wise F1 Score (Support)}}                                                                                                                                                                                                                                                                                                                                                                                                     \\ 
\cline{3-8}
                                  &                                             & \begin{tabular}[c]{@{}c@{}}\textbf{Audio Record}\\\textbf{(5)}\end{tabular} & \begin{tabular}[c]{@{}c@{}}\textbf{BrowserInfo}\\\textbf{(13)}\end{tabular} & \begin{tabular}[c]{@{}c@{}}\textbf{Contacts}\\\textbf{(2,343)}\end{tabular} & \begin{tabular}[c]{@{}c@{}}\textbf{GPS}\\\textbf{(522)}\end{tabular} & \begin{tabular}[c]{@{}c@{}}\textbf{Photos}\\\textbf{(662)}\end{tabular} & \begin{tabular}[c]{@{}c@{}}\textbf{URL}\\\textbf{(91)}\end{tabular}  \\ 
\hline
Extra Trees                       & 0.82                                        & 0.44                                                                        & 0.58                                                                        & 0.99                                                                        & 0.99                                                                 & 0.99                                                                    & 0.90                                                                 \\ 
\hline
\textbf{XGBoost}                  & \textbf{0.83}                               & \textbf{0.44}                                                               & \textbf{0.64}                                                               & \textbf{0.99}                                                               & \textbf{0.98}                                                        & \textbf{0.99}                                                           & \textbf{0.89}                                                        \\ 
\hline
K-nearest neighbors               & 0.79                                        & 0.40                                                                        & 0.50                                                                        & 0.99                                                                        & 0.98                                                                 & 0.99                                                                    & 0.86                                                                 \\
\hline
\end{tabular}
\end{adjustbox}
\end{table}


\vspace{-10mm}

\subsection{Target Classification}
Table \ref{targetDetectionTable} illustrates the results for classifying the target of the malicious actions predicted by the first stage. Due to the non-linearity posed by the datastream we consider tree-based algorithms such as ExtraTrees and XGBoost. Although XGBoost and ExtraTrees display comparable performances we prefer XGBoost to be integrated with our final pipeline since it has proven to be more scalable than the latter and displays the highest average performance of the models considered for the second stage. 




With less than 9\% inaccuracy in detecting malware from the first stage, we can predict with 83\% certainty on what kind of data is being stolen when we use an Extra trees  detector (Table \ref{malwareDetectionResults}) coupled with an XGBoost classifier (Table \ref{targetDetectionTable}).

\begin{figure}
    \centering
    \includegraphics[width=1\columnwidth,]{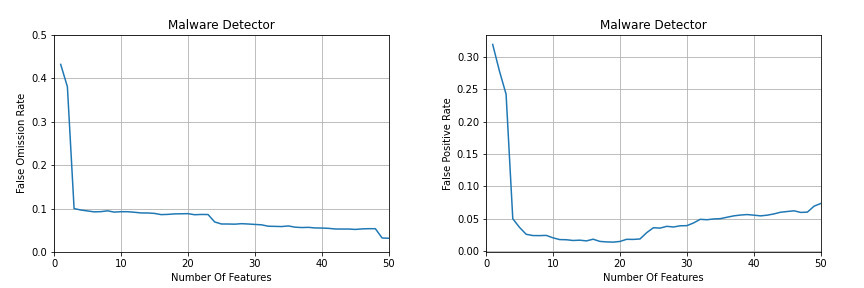}   \caption{Stepwise Forward Selection Convergence - Malware Detection}
    \label{stage_1_ffs}
\vspace{-5mm}
\end{figure}
\begin{figure}
    \centering
    \includegraphics[scale = 0.5]{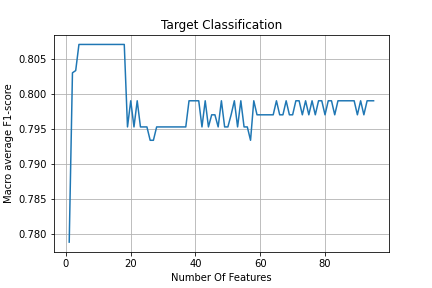}
    \caption{Stepwise Forward Selection Convergence - Target Classification}
    \label{stage_2_ffs}
\end{figure}

Furthermore, by using our feature selection approach we maintain the aforementioned model performance with the feature set reduced to approximately 0.1\% of the original set. 
Stepwise forward selection for malware detection (illustrated in Fig. \ref{stage_1_ffs}) reveals that we only require 10 features to determine if an action is malicious to achieve a minimum FOR and FPR of 0.087 and 0.019 respectively. Fig. \ref{stage_2_ffs} illustrates stepwise forward selection for target classification and suggests that we require only 16 features to categorize the type of data stolen. As a result of using such a small feature set, we minimize our throughput and processing time drastically.

\newpage
\section{Conclusion}
\label{conclusion}
In this paper, we propose and successfully test a two-stage machine learning model on the SherLock dataset to detect malicious actions in a smartphone and identify the type of data it steals. We successfully reduce one of the largest datasets for malware classification (SherLock) to 0.1\% of its initial feature set using our data preprocessing techniques and feature selection techniques. Furthermore, we go on to propose using False omission rate and False Positive Rate in conjunction to evaluate malware detectors. With 8.7\% inaccuracy in detecting malware from the first stage, our model can predict with 83\% certainty on what kind of data is being stolen when we use an Extra trees detector coupled with a XGBoost classifier. We exhibit our detector's robustness with the gradual decrease in the required train data from one user to achieve the aforementioned performance by training on more users and data. Anomaly/Outlier detection techniques for malware fail, since malicious actions do not lie outside benign distributions as conventionally expected.

\section{Future Works}
Although the proposed model reduces the percentage of train data required by a user to the minimum, malware detection is still dependent on user behaviour to work. There exists a need for a truly user-independent machine learning solution for malware detection to enhance user experience and ergonomics. 

\section{Acknowledgments}
We would like to thank all the talented members at Solarillion Foundation without whom this work would have taken much longer to carry out.

\vspace{12pt}


\begin{thebibliography}{00}
\bibitem{mcAfee2020} Raj Samani, "McAfee Mobile Threat Report Q1", 2020, URL: https://www.mcafee.com/content/dam/consumer/en-us/docs/2020-Mobile-Threat-Report.pdf
\bibitem{usSEC2015} U.S. Securities and Exchange Commission, "The Need for Greater Focus on the Cybersecurity Challenges Facing Small and Midsize Businesses", 2015, URL: https://www.sec.gov/news/statement/cybersecurity-challenges-for-small-midsize-businesses.html 
\bibitem{ibm2020}IBM, Cost of a Data Breach Report 2020, URL: https://www.ibm.com/security/digital-assets/cost-data-breach-report/
\bibitem{ShmmidtExecutables} Schmidt, A.-D., Bye, R., Schmidt, H.-G., Clausen, J., Kiraz, O., Yuksel, K. A., … Albayrak, S. (2009). Static Analysis of Executables for Collaborative Malware Detection on Android. 2009 IEEE International Conference on Communications.
\bibitem{demystified}  A. P. Felt, E. Chin, S. Hanna, D. Song, and D. Wagner,“Android Permissions Demystied,” in Proceedings of the 18th ACM Conference on Computer and Communications Security, 2011.
\bibitem{taintdroid} Enck, W., Gilbert, P., Chun, B. G., Cox, L. P., Jung, J., McDaniel, P., \& Sheth, A. N. (2019). TaintDroid: An information-flow tracking system for realtime privacy monitoring on smartphones. In Proceedings of the 9th USENIX Symposium on Operating Systems Design and Implementation, OSDI 2010.
\bibitem{sandbox} T. Bläsing, L. Batyuk, A. Schmidt, S. A. Camtepe and S. Albayrak, "An Android Application Sandbox system for suspicious software detection," 2010 5th International Conference on Malicious and Unwanted Software, Nancy, Lorraine, 2010.
\bibitem{andromaly} A. Shabtai, U. Kanonov, Y. Elovici, C. Glezer, Y. Weiss, “Andromaly: a behavioral malware detection framework for Android devices,” Journal of Intelligent Information Systems, vol. 38, no. 1, pp. 161-190, 2012.
\bibitem{limitsofstatic} A. Moser, C. Kruegel and E. Kirda, "Limits of Static Analysis for Malware Detection," Twenty-Third Annual Computer Security Applications Conference (ACSAC 2007).
\bibitem{microsoft2018} Royi Ronen, Marian Radu, Corina Feuerstein, Elad Yom-Tov, Mansour Ahmadi, "Microsoft Malware Classification Challenge", 2018, https://arxiv.org/abs/1802.10135
\bibitem{trafficAV}Shanshan Wang, Chen, Z., Zhang, L., Yan, Q., Yang, B., Peng, L., \& Zhongtian Jia. (2016). TrafficAV: An effective and explainable detection of mobile malware behavior using network traffic. 2016 IEEE/ACM 24th International Symposium on Quality of Service (IWQoS).
\bibitem{highImbalanceNetwork} Chen, Zhenxiang \& Yan, Qiben \& Han, Hongbo \& Wang, Shanshan \& Peng, Lizhi \& Wang, Lin \& Yang, Bo. (2017). Machine Learning Based Mobile Malware Detection Using Highly Imbalanced Network Traffic. Information Sciences. 433-434.
\bibitem{arora} A. Arora, S. Garg and S. K. Peddoju, "Malware Detection Using Network Traffic Analysis in Android Based Mobile Devices," 2014 Eighth International Conference on Next Generation Mobile Apps, Services and Technologies, Oxford, 2014, pp. 66-71. 
\bibitem{unkownMalware} D. Bekerman, B. Shapira, L. Rokach and A. Bar, "Unknown malware detection using network traffic classification," 2015 IEEE Conference on Communications and Network Security (CNS), Florence, 2015, pp. 134-142.
\bibitem{Sherlock} Y. Mirsky, A. Shabtai, L. Rokach, B. Shapira, and Y. Elovici, “Sherlock vs Moriarty: A Smartphone Dataset for Cybersecurity Research,” in Proceedings of the 2016 ACM Workshop on Artificial Intelligence and Security.
\bibitem{mobileAppAndMalwareClassification} Zheng, Yong \& Srinivasan, Sridhar. (2020). Mobile App and Malware Classifications by Mobile Usage with Time Dynamics. 
\bibitem{bigmomal} Sarah Wassermann and Pedro Casas. 2018. BIGMOMAL: Big Data Analytics for Mobile Malware Detection. In Proceedings of the 2018 Workshop on Traffic Measurements for Cybersecurity (WTMC '18). Association for Computing Machinery, New York, NY, USA, 33–39.
\bibitem{apache} Memon, Laraib \& Bawany, Narmeen \& Shamsi, Jawwad. (2019). A COMPARISON OF MACHINE LEARNING TECHNIQUES FOR ANDROID MALWARE DETECTION USING APACHE SPARK. Journal of Engineering Science and Technology. 
\bibitem{linuxMan}Linux Manual : https://man7.org/linux/man-pages/man5/proc.5.html
\bibitem{lightgbm} Guolin Ke, Qi Meng, Thomas Finley, Taifeng Wang, Wei Chen, Weidong Ma, Qiwei Ye, Tie-Yan Liu. "LightGBM: A Highly Efficient Gradient Boosting Decision Tree". Advances in Neural Information Processing Systems 30 (NIPS 2017), pp. 3149-3157.
\bibitem{forwardFeatureSelection} Han, J., Kamber, M., \& Pei, J. (2012). Data Preprocessing. Data Mining, 83–124. doi:10.1016/b978-0-12-381479-1.00003-4 
\bibitem{biology}  Cheng Chen, Qingmei Zhang, Qin Ma, Bin Yu, LightGBM-PPI Cheng Chen, Qingmei Zhang, Qin Ma, Bin Yu, LightGBM-PPI: Predicting protein-protein interactions through LightGBM with multi-information fusion, Chemometrics and Intelligent Laboratory Systems, Volume 191, 2019, Pages 54-64, ISSN 0169-7439.
\end{thebibliography}
\end{document}